\documentclass[10pt,aps,prl,twocolumn,groupedaddress,showkeys]{revtex4-1}
\usepackage{amsmath}
\usepackage{graphicx}
\DeclareGraphicsExtensions{.pdf,.eps,.png,.jpg,.mps}
\usepackage{epstopdf}
\usepackage{threeparttable}
\usepackage{color}
\begin{document}
\title{High-mobility, wet-transferred graphene grown by chemical vapor deposition}
\author{D. De Fazio$^{{1} (\ddagger)}$, D. G. Purdie$^{1 (\ddagger)}$, A. K. Ott$^1$, P. Braeuninger-Weimer$^2$, T. Khodkov$^3$, S. Goossens$^3$, T. Taniguchi$^4$, K. Watanabe$^4$, P. Livreri$^5$, F. H. L. Koppens$^3$, S. Hofmann$^2$, I. Goykhman$^1$, A. C. Ferrari$^1$, A. Lombardo$^1$}
\email[]{al515@cam.ac.uk}
\affiliation{$^1$ Cambridge Graphene Centre, University of Cambridge, Cambridge CB3 0FA, UK}
\affiliation{$^2$ Department of Engineering, University of Cambridge, 9 J.J. Thomson Avenue, CB3 0FA, Cambridge, UK}
\affiliation{$^3$ ICFO-Institut de Ciencies Fotoniques, The Barcelona Institute of Science and Technology, 08860 Castelldefels (Barcelona), Spain}
\affiliation{$^4$ National Institute for Materials Science, 1-1 Namiki, Tsukuba 305-0044, Japan}
\affiliation{$^5$ Universita' degli Studi di Palermo, 90128 Palermo, Italy}
\affiliation{($\ddagger$) equal contribution}
\begin{abstract}
We report high room-temperature mobility in single layer graphene grown by Chemical Vapor Deposition (CVD) after wet transfer on SiO$_2$ and hexagonal boron nitride (hBN) encapsulation. By removing contaminations trapped at the interfaces between single-crystal graphene and hBN, we achieve mobilities up to$\sim70000cm^2 V^{-1} s^{-1}$ at room temperature and$\sim120000cm^2 V^{-1} s^{-1}$ at 9K. These are over twice those of previous wet transferred graphene and comparable to samples prepared by dry transfer. We also investigate the combined approach of thermal annealing and encapsulation in polycrystalline graphene, achieving room temperature mobilities$\sim30000 cm^2 V^{-1} s^{-1}$. These results show that, with appropriate encapsulation and cleaning, room temperature mobilities well above 10000$cm^2 V^{-1} s^{-1}$ can be obtained in samples grown by CVD and transferred using a conventional, easily scalable PMMA-based wet approach.
\end{abstract}
\maketitle
\section{\label{In}Introduction}
Charge carriers in single layer graphene (SLG) can reach room-temperature (RT) mobility ($\mu$)$>$200000$cm^2V^{-1}s^{-1}$ at technologically-relevant carrier densities $n\sim10^{12}$cm$^{-2}$, corresponding to a sheet conductivity $\sigma\sim30mS$, with $\sigma=e\mu n$ and $e$ the elementary charge $1.6\times10^{-19}$C\cite{Sze2006}, limited by electron-phonon interactions\cite{ChenNN2008,ParkNL14}. High $\mu>$10000$cm^2V^{-1}s^{-1}$ at such carrier densities is essential for (opto)electronic devices\cite{RomaNRM3}, such as microwave transistors\cite{LinS2010}, photodetectors\cite{KoppNN2014}, THz detectors\cite{VicaNM11} and optical modulators\cite{SoriNP2018,RomaNRM3}. RT $\mu>100000cm^2V^{-1}s^{-1}$ can be achieved in micro-mechanically-cleaved (MC) samples, either suspended\cite{BoloSSC2008,DuNN2008,BoloN2009,MayoNL2012} or encapsulated in hexagonal boron nitride (hBN)\cite{WangS2013,MayoNL2011,PurdNC2018}. This is twice that of InSb\cite{BennSSE2005,OrrPRB2008,WangS2013} and InAs\cite{BennSSE2005,WangS2013} at $n$ up to$\sim4.5\times10^{12}cm^{-2}$ (corresponding to Fermi level  $E_F\sim270meV$)\cite{ChenNN2008,WangS2013}. Integration of SLG in a foundry requires scalable production and fabrication methods, to meet the requirements for 300mm wafers\cite{RomaNRM3}.
Chemical vapor deposition (CVD) allows production of SLG with lateral sizes up to hundreds of meters\cite{KobaAPL102}. A Cu foil is widely used as substrate due to its low carbon solubility ($\sim0.005$ carbon weight \% at $1084^{\circ}$C)\cite{LopeSM2004} and its catalytic role during growth\cite{LiNL2009,MattJMC2011}. Polycrystalline continuous films\cite{LiS2009,BaeNN2010} or isolated single crystals\cite{YuNM2011,LiJACS2011,GaoNC2012,Vaid2DM2017} can be grown by tuning parameters such as partial pressures\cite{VlasACS2011}, temperature $T$\cite{BhavNL2010} or substrate roughness\cite{YanACS2012,WangJACS2012}.

In order to be integrated into devices, SLG needs then to be removed from Cu and placed onto the target substrate. Several transfer methods have been developed, classified either as ``wet" or ``dry". We consider ``wet" all techniques whereby the SLG surface gets in contact with liquids, such water\cite{BonaMT2012}, solvents\cite{LiS2009,BaeNN2010} or other chemicals used to remove the substrate\cite{BonaMT2012} at any step of the transfer process. Wet transfer exploits a sacrificial layer (either a polymer\cite{YuAPL2008,LiS2009,BaeNN2010,WangACS2011,GannAPL2011,PetrNL2012,CalaAPL2014} or a thermal release tape\cite{BaeNN2010}) as support for SLG while the substrate is removed by chemical etching\cite{YuAPL2008,LiS2009,BaeNN2010}, electrochemical delamination\cite{WangACS2011} or selective interface etching\cite{WangAMI2016}. Wet transfer is simple and easily scalable\cite{BaeNN2010}, however it introduces polymer residuals or defects which typically reduce the SLG quality, resulting in undesired doping\cite{GannAPL2011} or low $\mu$ ($<3000cm^2V^{-1}s^{-1}$\cite{KobaAPL102}). The highest $\mu$ at RT reported to date in wet-transferred CVD SLG is$\sim20000cm^2 V^{-1}s^{-1}$\cite{GannAPL2011,CalaAPL2014}, achieved by placing SLG onto mechanically exfoliated hBN\cite{GannAPL2011,PetrNL2012,CalaAPL2014}. This is almost one order of magnitude lower than the state of the art in MC samples transferred by stamping\cite{PurdNC2018}.

Dry transfer consists in peeling SLG off the substrate without any chemical etching or electrochemical delamination. This exploits the Van der Waals interaction between SLG and hBN\cite{BansSA2015}. An hBN flake placed on a polymeric stamp is used to pick SLG\cite{BansSA2015}. The hBN-SLG stack is subsequently stamped onto a second hBN layer, achieving full encapsulation ($\emph{i.e.}$ the entire SLG area is enclosed between two hBN layers), so that SLG is never in contact with any liquid\cite{BansSA2015}. $\mu$ up to$\sim3\times10^6cm^2V^{-1}s^{-1}$ at 1.8K was reported for dry transferred, encapsulated, CVD grown SLG\cite{BansSA2015}, comparable to MC SLG encapsulated in hBN\cite{WangS2013,MayoNL2011,PurdNC2018}. Dry transfer requires Cu oxidation below the SLG surface\cite{BansSA2015}, to weaken the SLG interaction with Cu\cite{BansSA2015}. However, Cu oxidation can be a time-consuming process (a few days are required for oxidation of Cu underneath SLG islands width of few hundred $\mu$m at ambient conditions\cite{DroeAPL2017}), and its speed depends on the SLG coverage of the Cu foil, as SLG slows oxidation\cite{ChenACS2011}.

We reported a method to clean interfaces in heterostructures consisting of MC SLG and hBN\cite{PurdNC2018}, achieving atomically-clean interfaces and RT $\mu$ up to$\sim150000cm^2V^{-1}s^{-1}$ even in samples intentionally contaminated with polymers and solvents\cite{PurdNC2018}. Here we apply the same approach to CVD-grown single crystal (SC) SLG domains (lateral size $\sim 500 \mu$m) wet transferred on Si+SiO$_2$. We achieve $\mu$ up to$\sim70000cm^2V^{-1}s^{-1}$ at RT and$\sim120000cm^2V^{-1}s^{-1}$ at 9K, with ballistic transport over$\sim$600nm at 9K. To the best of our knowledge, this is the highest $\mu$ reported thus far in wet-transferred, CVD-grown SLG. We also apply the same approach to polycrystalline SLG (poly-SLG) with average grain size of few $\mu$m$^2$, achieving RT $\mu\sim7000cm^2V^{-1}s^{-1}$. By then annealing in Ar/H$_2$ at 600$^{\circ}$C we get RT $\mu$ up to$\sim30000cm^2V^{-1}s^{-1}$. Thus, by combining encapsulation and interface cleaning, both SC and poly CVD-grown SLG have RT $\mu\gg10000cm^2 V^{-1} s^{-1}$ even after exposure to contaminants such as etchants, polymers and solvents. Therefore scalable processing methods, such as polymer-based wet transfer, can be used for fabrication of (opto)electronic devices achieving the $\mu$ required for them to surpass existing technologies\cite{RomaNRM3}.
\begin{figure}
\centerline{\includegraphics[width=90mm]{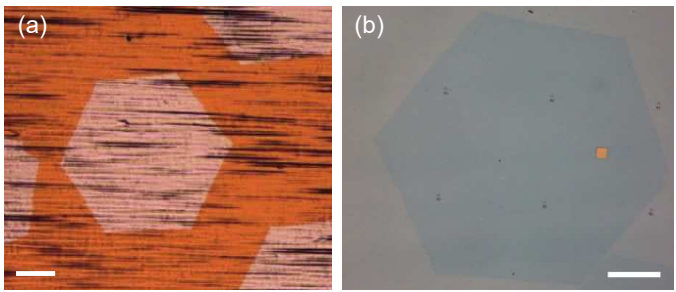}}
\caption{a) Hexagon-shaped SC-SLG on Cu. Contrast for imaging SC by optical microscopy is enhanced by heating in atmosphere at 250$^{\circ}$C for 1 min to promote Cu oxidation. b) SC-SLG wet transferred on SiO$_2$/Si with patterned lithography markers. Scale bars: $100\mu\text{m}$.}
\label{fig:Fig1}
\end{figure}
\section{\label{Disc}Results and Discussion}
SLG single crystals are grown on 25$\mu$m-thick Alfa Aesar, 99.8\% purity Cu foil, Fig.\ref{fig:Fig1}a. As for Ref.\citenum{BraeCM2016}, before growth the Cu foil is floated on the surface of a 30\% hydrogen peroxide (H$_2$O$_2$) solution at 100$^{\circ}$C for 5min to oxidize the back-side. This treatment favours the growth of single crystals due to oxygen-enabled scavanging of carbon impurities trapped in the Cu top surface and bulk, which would otherwise act as nucleation sites during the growth\cite{BraeCM2016}. Growth is then performed in a Aixtron Black Magic Pro, 4-inch cold wall plasma enhanced-CVD (PECVD) system with a base pressure$\sim$0.05mbar.  The foil is initially placed in the furnace and heated to$\sim$1065$^{\circ}$C in a Ar environment (200sccm) at 100$^{\circ}$C/min. Once the growth $T$ is reached, the Cu foil is annealed keeping $T$ constant in Ar (200sccm) for 30min. Carbon deposition is done in an Ar/H$_2$ environment (250/26sccm) using 9sccm CH$_4$, 0.1\% diluted in Ar, for 45min. Samples are then cooled in 250sccm Ar to RT. The crystallinity of the as grown-SLG is confirmed via transmission electron microscopy (TEM)\cite{BraeCM2016}. Electron diffraction on multiple spots reveals the same crystal orientation across the hexagonal SLG domain with$\sim500\mu m$ lateral size\cite{BraeCM2016}. SLG crystals are then transferred on Si+285nm SiO$_2$ using a wet method\cite{LiS2009,BaeNN2010,BonaMT2012}. A poly(methyl methacrylate) (PMMA) support layer is spin coated at the SLG surface, following which the samples are placed in a solution of ammonium persulfate (APS) and deionized (DI) water, whereby Cu is chemically etched\cite{LiS2009,BaeNN2010,BonaMT2012}. The PMMA/SLG stack is then moved to a beaker with DI water to remove APS residuals and lifted with the target SiO$_2$/Si substrate. After drying, PMMA is removed in acetone, leaving SLG on SiO$_2$/Si, Fig.\ref{fig:Fig1}b. We then encapsulate SLG in hBN. hBN bulk crystals are grown as for Ref.\citenum{TaniJCG2007}. These are mechanically exfoliated on Si+285nm SiO$_2$ to be used for SLG encapsulation. Flakes of lateral size$>100\mu$m are identified and selected by inspecting the SiO$_2$/Si surface using a combination of bright and dark field optical microscopy, Raman spectroscopy and atomic force microscopy (AFM). For the top hBN we use flakes with thickness t$_{hBN}\sim$2-300nm. For the bottom hBN, flakes with $t_{hBN}>$10nm are chosen, as thinner ones do not screen roughness and charged impurities of the underlying SiO$_2$\cite{DeanNN2010,PurdNC2018}.
\begin{figure*}
\centerline{\includegraphics[width=190mm]{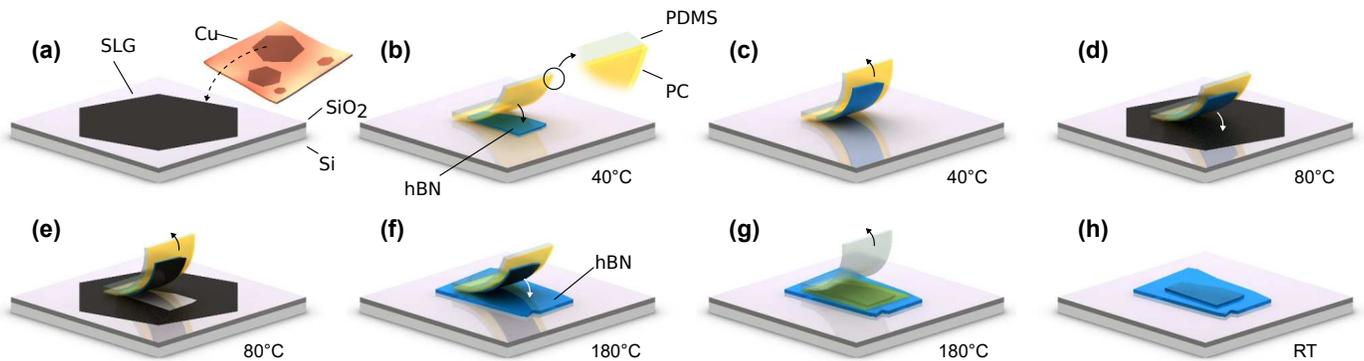}}
\caption{Transfer process to encapsulate SLG in hBN.(a) SLG (black hexagon) is wet-transferred from Cu to Si+285nm SiO$_{2}$. In parallel, (b-c) top hBN (light blue) is picked up with a PC/PDMS stamp (yellow+white). (d-e) SLG is then picked up with the bottom hBN from Si/SiO$_2$ at 80$^{\circ}$C and (f-g) hot-released at 180$^{\circ}$C on the bottom hBN, (h) leaving the encapsuled SLG}
\label{fig:Fig2}
\end{figure*}
\begin{figure*}
\centerline{\includegraphics[width=190mm]{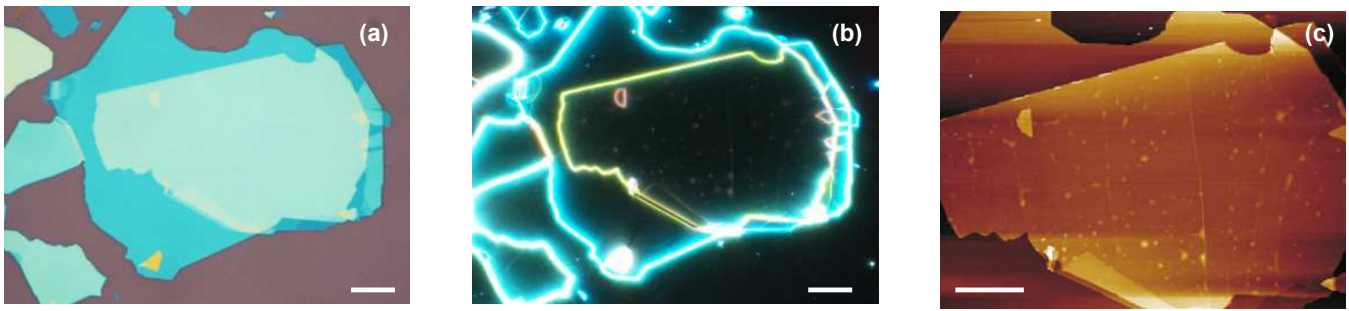}}
\caption{a) Bright, b) dark field and c) AFM images of SC-SLG encapsulated in hBN. Scale bars 10$\mu\text{m}$}
\label{fig:Fig3}
\end{figure*}

The encapsulation of SLG in hBN typically results in blisters containing trapped adsorbates and contaminants\cite{KretNL2014}, which must be avoided as they locally degrade transport\cite{KretNL2014}. Ref.\citenum{PizzNC2016} showed how to remove contamination blisters by a hot pick-up technique. This uses $T$ above the glass transition, $T_{g}$, of the polymer stamp during encapsulation, allowing the interfaces of two materials to be brought together in a directional, conformal manner\cite{PizzNC2016}. We modified this approach by using polycarbonate (PC) stamps at $T=180^{\circ}$C, achieving fast ($>10\mu m/s$) removal of contaminants in fully-encapsulated hBN/SLG/hBN heterostructures\cite{PurdNC2018}. This results in atomically-flat interfaces even on samples intentionally contaminated with polymers and solvents, achieving in all cases RT $\mu>150000cm^2V^{-1}s^{-1}$\cite{PurdNC2018}. As for Ref.\citenum{PurdNC2018}, here we use a transfer stamp consisting of a PC film on a polydimethylsiloxane (PDMS) block for mechanical support, placed on a glass slide attached to a micro-manipulator, enabling fine ($\sim1\mu m$) spatial control in x, y and z. $T$ is set using a heated stage. The process is depicted in Fig.\ref{fig:Fig2}. It begins by positioning the stamp above a hBN flake, then lowering it into contact, with the stage $T$ set to $40^{\circ}$C. As the stamp is withdrawn, the hBN adheres to the PC surface, and is delaminated from SiO$_{2}$, Fig.\ref{fig:Fig2}b,c. The picked-up hBN is then positioned above the (wet transferred) SLG on SiO$_2$ and brought into contact at $80^{\circ}$C. We then wait$\sim$5mins to promote adhesion between hBN and SLG, after which the stamp is lifted, picking up the SLG portion in contact with hBN, Fig.\ref{fig:Fig2}d,e. The final step consists in bringing the top hBN/SLG into contact with the bottom hBN at $180^{\circ}$C, with the stamp tilted to ensure that contact occurs first on one side, then conformally advancing across the substrate, Fig.\ref{fig:Fig2}f. At $180^{\circ}$C, withdrawing the stamp releases the PC onto the substrate, Fig.\ref{fig:Fig2}g, then dissolved by placing the sample in chloroform, Fig.\ref{fig:Fig2}h. Figs.\ref{fig:Fig3}a,b are bright and dark field images of the final hBN/SLG/hBN stack. Scans by AFM, Fig.\ref{fig:Fig3}c, reveal that some blisters are present. They tend to aggregate in specific areas, mostly along vertical lines, which we attribute to residual wrinkles from wet transfer.
\begin{figure*}
\centerline{\includegraphics[width=180mm]{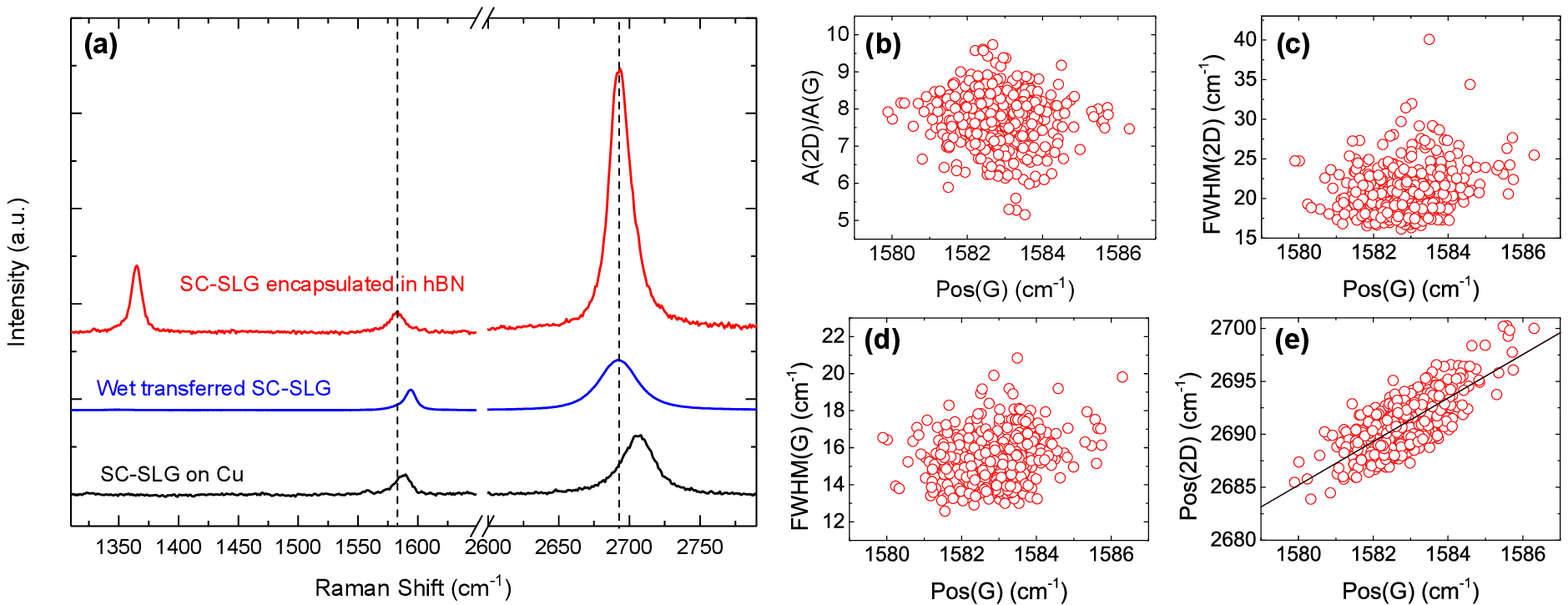}}
\caption{a) Raman spectra of SC-SLG on Cu (black), transferred on SiO$_2$ (blue) and after encapsulation (red). In the spectrum on Cu, PL of Cu is subtracted. b,c,d,e) Plots of b) A(2D)/A(G), c) FWHM(2D), d) FWHM(G), e) Pos(2D) a function of Pos(G) mapped across a $20\mu m\times30\mu m$ region in an encapsulated SC-SLG}
\label{fig:Fig4}
\end{figure*}

The SLG quality is monitored at each step of the fabrication process by Raman spectroscopy. Raman measurements are performed with a Renishaw InVia spectrometer equipped with a 100$\times$ objective, 2400l/mm grating at 514nm. The power on the sample is below$\sim$1mW to avoid any heating and damage. Fig.\ref{fig:Fig4}a plots representative spectra of a SC-SLG on Cu (black), after transfer on SiO$_2$/Si (blue) and encapsulated in hBN (red). The SLG spectrum on Cu is shown after subtraction of the Cu photoluminescence (PL)\cite{LagaAPL2013}. This has a 2D peak with a single Lorentzian shape and a with full width at half maximum FWHM(2D)$\sim$27cm$^{-1}$, a signature of SLG\cite{FerrPRL2006}. The position of the G peak, Pos(G), is$\sim$1588cm$^{-1}$, with FWHM(G)$\sim$14cm$^{-1}$. The 2D peak position, Pos(2D) is$\sim$2706cm$^{-1}$, while the 2D to G peak intensity and area ratios, I(2D)/I(G) and A(2D)/A(G), are$\sim$3 and$\sim$5.5. No D peak is observed, indicating negligible defects\cite{CancNL2011,FerrNN2013}. After wet transfer on SiO$_2$/Si, the 2D peak retains its single-Lorentzian lineshape with FWHM(2D)$\sim$30cm$^{-1}$. The D peak is still negligible, indicating that no significant defects are induced by wet transfer. Pos(G) is$\sim$1594cm$^{-1}$, FWHM(G)$\sim$11cm$^{-1}$, Pos(2D)$\sim$2692cm$^{-1}$, I(2D)/I(G) and A(2D)/A(G) are$\sim$2.5 and $\sim$6.8, indicating a doping$<$200meV\cite{DasNN2008,BaskPRB2009}. After pick up and encapsulation, we perform a Raman mapping over an area$\sim$20$\mu$m $\times 30\mu$m, using the same measurement conditions. A representative spectrum is shown in red in Fig.\ref{fig:Fig4}a. This comprises both the SLG signatures and the hBN $E_{2g}$ peak$\sim$1364cm$^{-1}$, with FWHM$\sim$9.5cm$^{-1}$, as expected for bulk hBN\cite{ArenNL2006,ReicPRB2005,NemaPRB1979}. The hBN E$_{2g}$ peak is a combination of top and bottom hBN and may overlap a small D peak.

Fig.\ref{fig:Fig4}b-e plot Raman data extracted from mapping: A(2D)/A(G), FWHM(2D), FWHM(G) and Pos(2D) as a function of Pos(G). Pos(G) depends on both doping\cite{DasNN2008,BaskPRB2009} and strain\cite{MohiPRB2009}. This implies that local variations in strain and doping manifest as a spread in Pos(G) which, in our sample, varies from 1580 to 1586cm$^{-1}$. The samples have narrow 2D peaks with mean FWHM(2D)$\sim$20$\pm$3cm$^{-1}$. Figs.\ref{fig:Fig4}b-d plot A(2D)/A(G) and FWHM(G) as function of Pos(G), indicating small variations of both doping and strain within the SLG layer. The mean A(2D)/A(G) is$\sim$7.8$\pm$0.7, while the mean FWHM(G) is$\sim$15.4$\pm$1.2cm$^{-1}$. The high A(2D)/A(G) is an indication of intrinsic samples\cite{DasNN2008}, when combined with FWHM(G)$\sim15$cm$^{-1}$, as well as Pos(G)$\sim1583$cm$^{-1}$ and Pos(2D)$\sim 2693$cm$^{-1}$\cite{DasNN2008}.
\begin{figure*}
\centerline{\includegraphics[width=180mm]{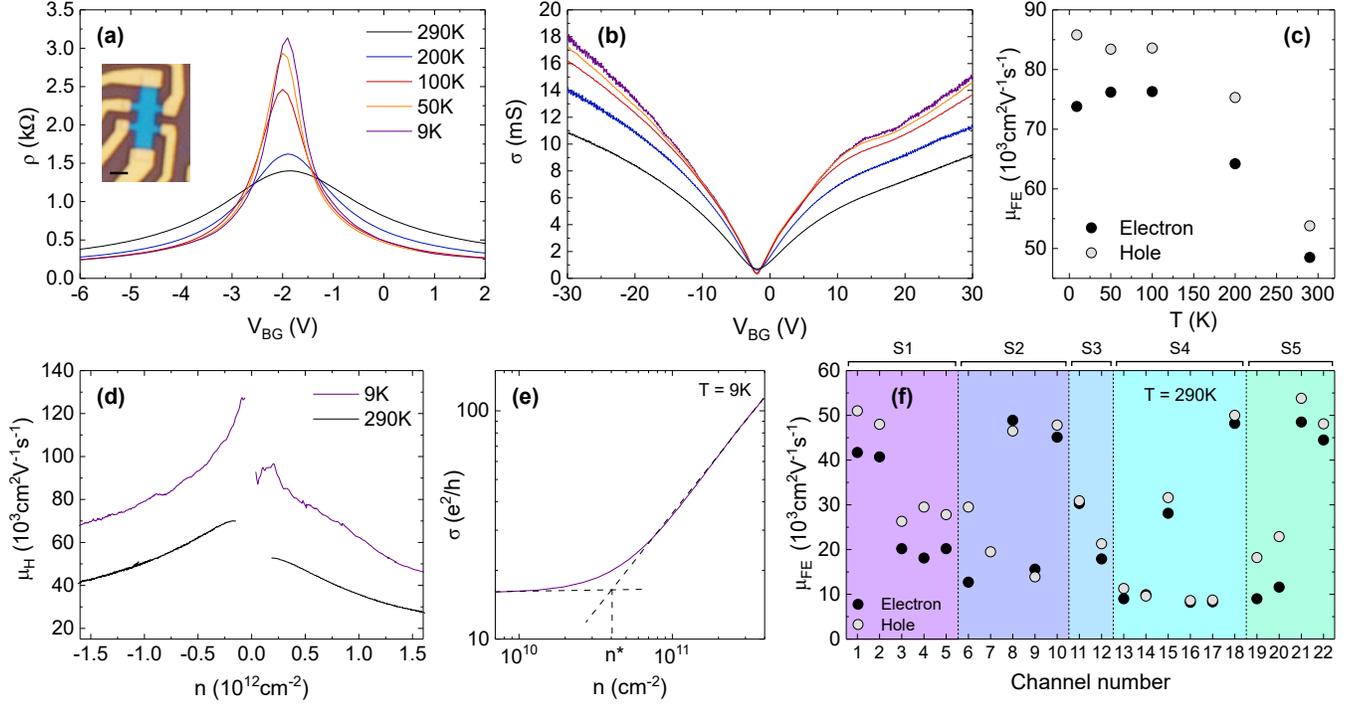}}
\caption{a) Resistivity as a function of gate voltage between 9 and 290K. Inset: Optical image of the device, with scale bar $1.5\mu\text{m}$. b) Conductivity as a function of gate voltage. c) Field effect mobility as a function of $T$. d) Density dependent $\mu$ from the Drude model $\mu=\sigma\slash ne$ at 9 and 290K. e) Disorder induced charge inhomogeneity by plotting conductivity as a function of $n$ close to the neutrality point on logarithmic axes. f) Field effect mobility from 22 Hall bars for 5 encapsulated SC-SLG. Each shaded region corresponds to a different sample (S1-S5), while black (grey) dots refer to electron (hole) mobilities in different channels within the same SC-SLG}
\label{fig:Fig5}
\end{figure*}

The rate of change of Pos(2D) and Pos(G) with strain is ruled by the Gr\"{u}neisen parameters ($\gamma$)\cite{MohiPRB2009}, which relate the relative change in the peak positions in response to strain($\epsilon$), $\emph{i.e.}$ $\gamma =[\omega - \omega_{0}]/[2\epsilon\omega_{0}]$, where $\omega$ is the frequency of Raman peaks at finite strain and $\omega_{0}$ the frequency at zero strain\cite{MohiPRB2009}. For biaxial strain the Gr\"{u}neisen parameters for G and 2D peak are respectively $\gamma_{G}\sim1.8$ and $\gamma_{2D}\sim2.6$, resulting in $\Delta$Pos(2D)/$\Delta$Pos(G)$\sim2.5$\cite{MohiPRB2009,ZabeNL2012,ProcPRB2009}. In the case of uniaxial strain $\gamma_{G}\sim1.8$\cite{MohiPRB2009}, however extraction of $\gamma_{2D}$ is not straightforward, as uniaxial strain shifts the relative position of the SLG Dirac cones\cite{MohiPRB2009,ZabeNL2012}, which in turn affects the 2D peak, as this is involves by intervalley scattering\cite{FerrNN2013,MohiPRB2009}. Ref.\cite{MohiPRB2009} experimentally derived an upper bound $\gamma_{2D}\sim3.55$ and theoretically calculated $\gamma_{2D}\sim2.7$, consistent with experimentally reported $\Delta$Pos(2D)/$\Delta$Pos(G)$\sim2-3$\cite{MohiPRB2009,YoonPRL2011,LeeNC2012}. Biaxial strain can be differentiated from uniaxial from the absence of G and 2D peak splitting with increasing strain\cite{FerrNN2013}, however at low ($\lesssim0.5\%$) strain the splitting cannot be resolved\cite{MohiPRB2009,YoonPRL2011}. Fig.\ref{fig:Fig4}e shows the correlation between Pos(2D) and Pos(G). A linear fit gives a slope $\Delta$Pos(2D)/$\Delta$Pos(G)$\sim$2.07. The presence (or coexistence) of biaxial strain cannot be ruled out. For uniaxial(biaxial) strain, Pos(G) shifts by $\Delta $Pos(G)$\slash\Delta \epsilon\sim 23(60) \text{cm}^{-1}\slash \%$\cite{MohiPRB2009,YoonPRL2011,ZabeNL2012}. For intrinsic SLG (E$_{F}\ll 100$meV), the unstrained, undoped Pos(G) is$\sim$1581.5cm$^{-1}$\cite{FerrPRL2006, PiscPRL2004}. In Fig.\ref{fig:Fig4}d the mean Pos(G) is 1582.8cm$^{-1}$, which would lead to uniaxial(biaxial) strain $\epsilon\sim$0.05\%(0.02\%)\cite{MohiPRB2009,YoonPRL2011,ZabeNL2012}. The scattering of A(2D)/A(G) within the mapped area indicates small ($\ll$100meV) variation of doping.
\begin{figure*}
\centerline{\includegraphics[width=180mm]{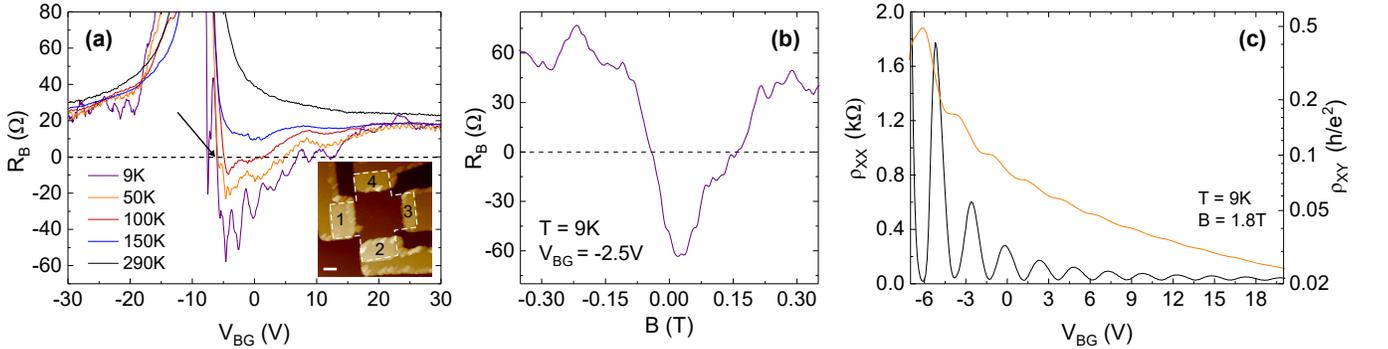}}
\caption{a) Bend resistance of a Hall cross with arm width 640nm as a function of V$_{BG}$ for $T$=9-290K. The negative resistance below 100K for positive V$_{BG}$ indicates that carriers travel ballistically from contacts 1 to 3. The arrow indicates $R_{B}=0$ at 100K, $\emph{i.e.}$ the transition from diffusive to ballistic transport at which $l_{m}\sim 640nm$. Inset: AFM scan of the sample with numbered contacts. Scale bar $200\text{nm}$. (b) Bend resistance as a function of magnetic field, at $V_{BG}=-2.5V$ and 9K. (c) $\rho_{xx}$ and $\rho_{xy}$ as a function of $V_{BG}$ at 9K and B=1.8T}
\label{fig:Fig6}
\end{figure*}

We then process our samples into Hall-bars, in order to perform 4-terminal electrical transport. To form contacts, we follow Ref.\citenum{WangS2013}. After defining the shape of the desired Hall bar using a hard mask\cite{PurdNC2018}, patterned by e-beam lithography, we remove the unmasked part by reactive ion etching (RIE) using a plasma formed from a mix of tetrafluoromethane ($\text{CF}_{4}$) and oxygen ($\text{O}_{2}$) (ratio 4:1) under a 20W forward radio frequency (RF) power. SLG is then contacted at its edges\cite{WangS2013} by depositing metal leads via e-beam evaporation of 5/70nm Cr/Au.

Fig.\ref{fig:Fig5}a plots the resistivity of encapsulated CVD SLG as a function of back gate voltage $V_{BG}$. An optical image of one of the Hall bars is in the inset. The corresponding conductivity is shown in Fig.\ref{fig:Fig5}b. We extract the field effect mobility $\mu_{FE}= C_{ox}^{-1}\left(d\sigma\slash dV_{BG}\right)$ by performing a linear fit to the conductivity close to the charge neutrality point (CNP)\cite{NovoS2004}, where $C_{ox}$ is the gate capacitance per unit area. This is calculated as$\sim$1.1$\cdot$10$^{-4}$F/m$^2$, assuming a parallel plate capacitor where the bottom hBN is in series with the 285nm SiO$_2$ layer\cite{PurdNC2018}. The bottom hBN thickness is$\sim$15nm for the sample in Fig.\ref{fig:Fig5}a. We assume a dielectric constant for hBN $\epsilon\sim$3, considering that values between 2-4 are usually reported\cite{KimACS2012}. Our $C_{ox}$ is orders of magnitude smaller than the quantum capacitance of SLG\cite{XiaNN2009}, which is therefore neglected. This yields $\mu_{FE}\sim49000cm^2V^{-1}s^{-1}$ and $54000cm^2V^{-1}s^{-1}$ for electrons and holes at 290K. $\mu_{FE}$ as a function of $T$ in Fig.\ref{fig:Fig5}c reaches$\sim86000cm^2V^{-1}s^{-1}$ for holes at 9K.

In order to extract $\mu$ as function of $n$, we investigate transport in the presence of an out of plane magnetic field B (Hall effect)\cite{Sze2006}. To distinguish $\mu$ calculated from Hall effect from that estimated from the slope of the conductivity, we indicate the first as $\mu_{H}$\cite{Sze2006}. Assuming a Drude model of conductivity $\mu_{H}=\sigma\slash ne$\cite{Sze2006}, where $n$ is extracted by measuring the Hall voltage with B=0.5T. $\mu_{H}$ in Fig.\ref{fig:Fig5}d reaches$\sim70000cm^2V^{-1}s^{-1}$ close to the CNP, while it is$>30000cm^2V^{-1}s^{-1}$ for $n$ up to$\sim1.5\cdot10^{12}\text{cm}^{-2}$. At 9K $\mu_{H}$ is$>120000cm^{2}V^{-1}s^{-1}$.

Such values are consistent with those reported in CVD-grown SLG encapsulated in hBN using dry techniques\cite{BansSA2015,BansNL2016}. \emph{E.g.}, Ref.\citenum{BansSA2015} measured in Hall bars made with CVD SLG encapsulated by dry transfer in hBN $\mu_{FE}\sim110000-350000cm^{2}V^{-1}s^{-1}$ at 1.6K. The wet transfer method in Ref.\cite{PetrNL2012} resulted in $\mu_{FE}$ up to$\sim50000cm^{2}V^{-1}s^{-1}$ at 1.6K\cite{PetrNL2012},$\sim2.5$ lower than here.
	
Fig.\ref{fig:Fig5}a shows that our SLG is nearly intrinsic, since the resistivity peak occurs close to zero gate voltage, at $V_{CNP}=-1.8V$, indicating moderate $n\sim1.2\cdot10^{11}\text{cm}^{-2}$, in agreement with the Raman analysis. Furthermore, the resistivity peak is narrow, with FWHM$\sim$1V at 9K. This places an upper bound on the disorder induced charge inhomogeneity, $n^{*}$, with narrower peaks corresponding to lower disorder\cite{CoutPRX2014,BoloSSC2008,KretNL2014}. $n^{*}<10^{11}\text{cm}^{-2}$ is typically associated with either suspended\cite{CoutPRX2014}, or dry encapsulated samples\cite{BansSA2015,PurdNC2018}. We extract $n^{*}$ for our sample as for Refs.\citenum{DuNN2008,CoutPRX2014}, Fig.\ref{fig:Fig5}e, and get$\sim 4\cdot10^{10}cm^{-2}$, one order of magnitude higher than clean MC samples\cite{PurdNC2018}.
	
In order to investigate the $\mu$ variation, we fabricate 22 Hall bars from 5 encapsulated SC-SLG. The distribution of $\mu_{FE}$ at RT in Fig.\ref{fig:Fig5}f ranges from a minimum$\sim10000cm^2V^{-1}s^{-1}$ up to$\sim55000cm^2 V^{-1}s^{-1}$. Bend resistance measurements as a function of $T$ are typically used to probe ballistic transport in SLG\cite{MayoNL2011,BansSA2015,WangS2013,PurdNC2018}, whereby current is injected around a bend while measuring the voltage. In SLG the carrier mean free path $l_{m}$ is related to conductivity\cite{HwanPRL2007}:
\begin{equation}
l_{m}=\frac{h}{2e^2}\sigma\sqrt{\frac{1}{\pi n}}
\label{eq:eqn1}
\end{equation}
where $h$ is the Planck constant. $l_{m}$ can be extracted from Fig.\ref{fig:Fig5}a. It varies from$\sim0.6\mu m$ at 290K up to$\sim1\mu m$ at 9K. For such $l_{m}$ we expect ballistic transport\cite{MayoNL2011}. We therefore fabricate a Hall cross with arm width $W=640\text{nm}$ to perform bend resistance measurements as a function of $T$\cite{MayoNL2011,WangS2013,BansNL2016}.
\begin{figure*}
\centerline{\includegraphics[width=180mm]{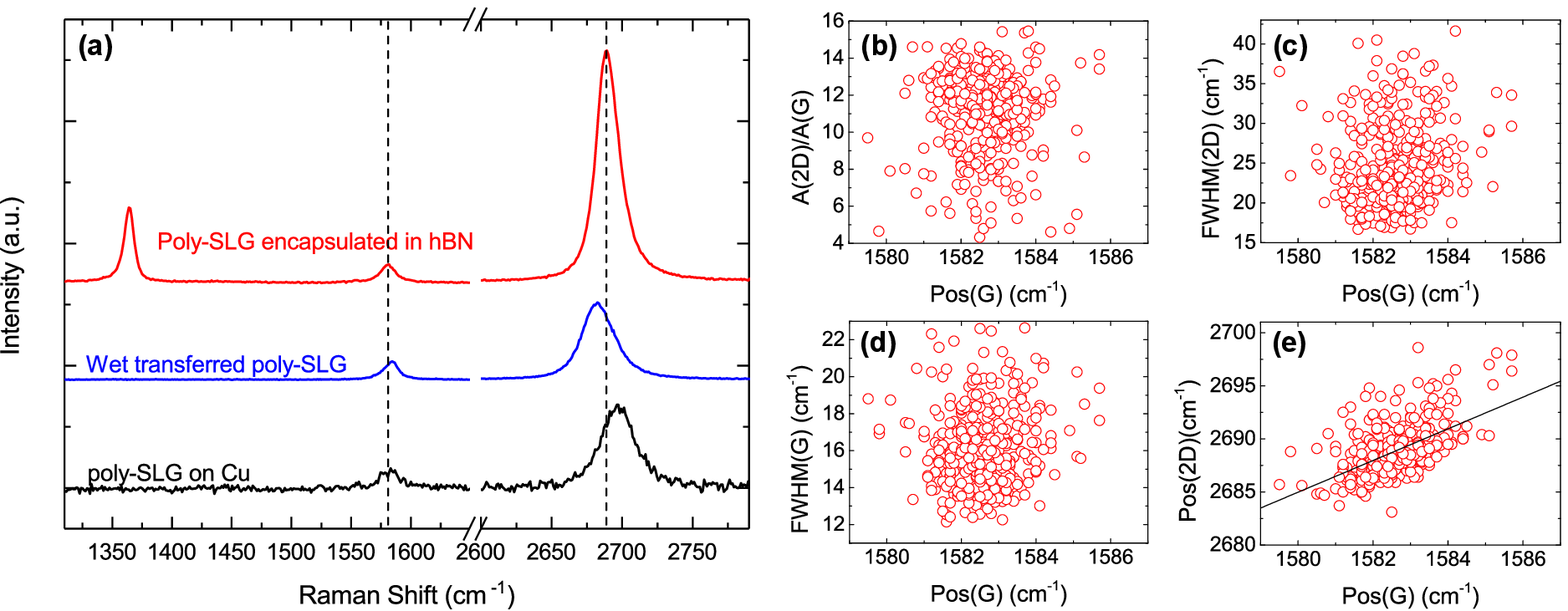}}
\caption{a) Raman spectra of Poly-SLG on Cu (black), transferred on SiO$_2$ (blue) and after encapsulation (red). In the spectrum on Cu, PL of Cu has been subtracted. b,c,d,e) plots of b) A(2D)/A(G), c) FWHM(2D), d) FWHM(G), e) Pos(2D) a function of Pos(G) mapped across a $18\mu m\times23\mu m$ region in an encapsulated Poly-SLG}
\label{fig:Fig7}
\end{figure*}
\begin{figure}
\includegraphics[width=90mm]{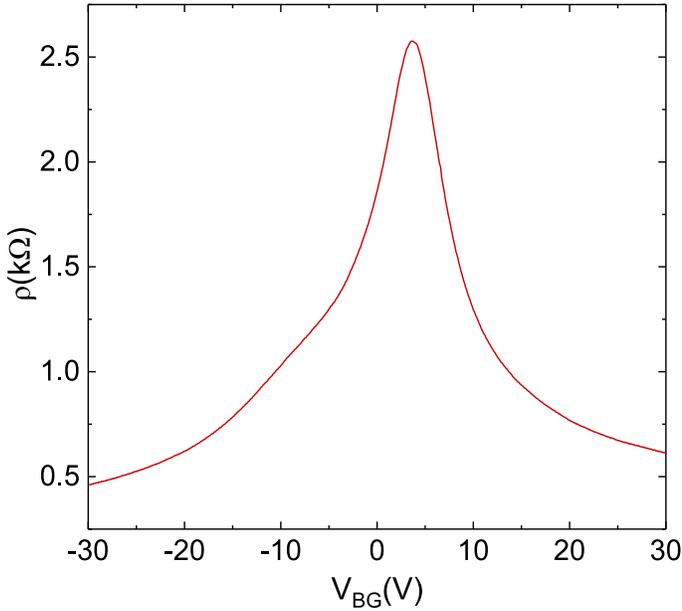}
\caption{Resistivity as a function of V$_{BG}$ of Poly-SLG in a Hall bar geometry at 290K.}
\label{fig:Fig8}
\end{figure}
	
An AFM scan of the Hall cross is in the inset of Fig.\ref{fig:Fig6}a. A constant current $I_{1,2}$ is passed between contacts 1 and 2, and the voltage between contacts 3 $(V_{3})$ and 4 $(V_{4})$ measured, from which $R_{B}=(V_{4}-V_{3})\slash I_{1,2}$ is extracted. In the diffusive regime, where $l_{m}<W$, a positive $R_{B}$ is measured, related to the sheet resistance via the Van-der Pauw formula\cite{VandPRR1958}: $\rho=\left(\pi \slash \ln2\right)R_{B}$. Alternatively, if $l_{m} \geq W$, carriers injected at contact 1 follow ballistic trajectories to contact 3, resulting in $R_{B}<0$. $R_{B}$ as a function of $T$ is in Fig.\ref{fig:Fig6}b. Below 100K, $R_{B}$ becomes negative for positive V$_{BG}$, indicating $l_{m} \geq 640\text{nm}$. At the transition from the diffusive to ballistic $R_{B}=0$ and $l_{m} \sim W$, from which $\mu$ may be estimated rearranging Eq.\ref{eq:eqn1} by substituting $l_m=W$, to give\cite{MayoNL2011}:
\begin{equation}
\mu \sim \frac{2e}{h}W\sqrt\frac{\pi}{n_{th}}
\label{eq:eqn2}
\end{equation}
where $n_{th}$ is the carrier concentration at which $R_{B}=0$. At 9K, $R_{B}=0$ occurs at $V_{BG}=-5.8V$, as shown by the arrow in Fig.\ref{fig:Fig6}a, corresponding to $\mu\sim129000cm^2V^{-1}s^{-1}$, in agreement with the peak $\mu_{H}$ from Fig.\ref{fig:Fig5}d. Further confirmation of ballistic transport is achieved by applying an out of plane B, which results in the bending of electrons from their straight-line trajectories, deflecting them away from contact 3, and switching $R_{B}$ from negative to positive, Fig.\ref{fig:Fig6}b. Magneto-transport at B=1.8T and 9K, Fig.\ref{fig:Fig6}c, shows resolved Landau levels, consistent with Refs.\citenum{PetrNL2012,BansSA2015}, proving the high electronic quality of our samples.
\begin{figure*}
\centerline{\includegraphics[width=180mm]{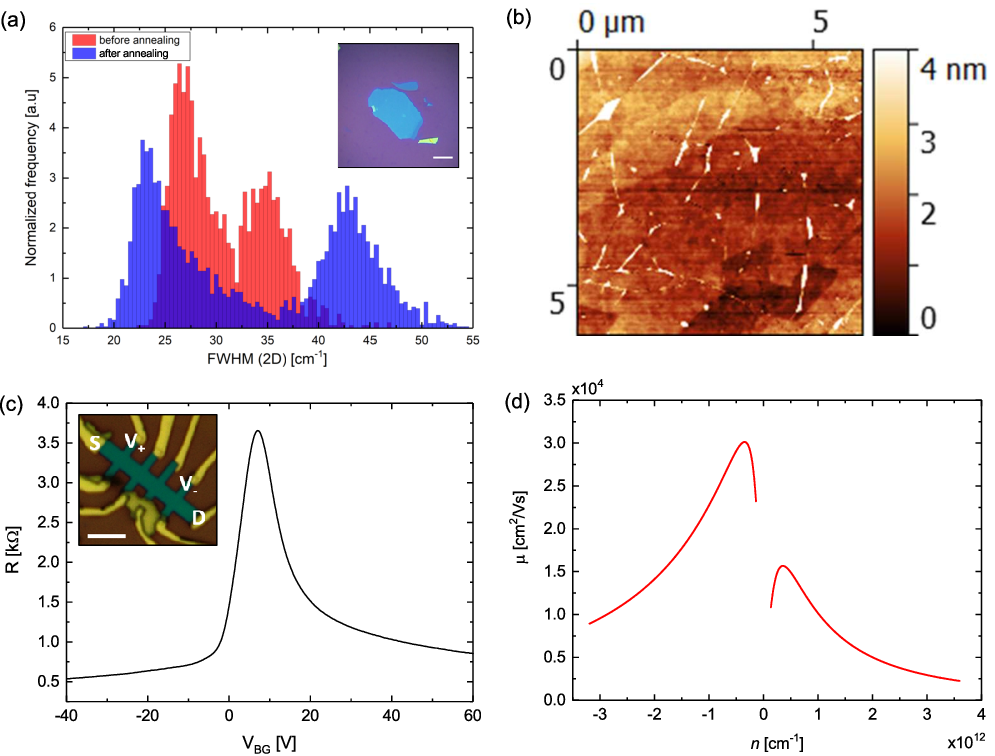}}
\caption{(a) Statistical distribution of FWHM(2D) for poly-SLG on hBN (blue) before and (red) after annealing in H$_2$/Ar at 600$^\circ$C. Spatially-resolved Raman spectra are acquired by mapping Poly-SLG layers partially on SiO$_2$ and partially encapsulated in hBN. The inset shows the sample mapped, scalebar 10 $\mu$m. b) AFM image of poly-SLG on hBN after annealing. c) four-terminal resistance of annealed and encapsulated poly-SLG as function of V$_{BG}$. (Inset) Optical picture of the device, scale bar 5$\mu$m. The bias is applied between (S) source and drain (D) electrodes and the voltage drop along the sample is sensed between V+ and V-. d) $\mu_{FE}$ as function of $n$, extracted by assuming a Drude model for the conductivity}
\label{fig:Fig9}
\end{figure*}
		
For all our samples $l_{m} \lesssim 1\mu \text{m}$ at 9K, even in Hall bars with channel widths$>3\mu\text{m}$. This is supported by the fact that Hall crosses with width $W=1\mu\text{m}$ exhibit no negative bend resistance at any $T$. Our maximum $l_{m}\sim 1\mu\text{m}$ is lower than dry transferred samples, in which $l_{m}$ typically exceeds the sample dimensions at low $T$\cite{WangS2013,BansNL2016}, resulting in edge scattering limited conductivity. Indeed, the highest reported $l_{m}$ in literature are$>10\mu\text{m}$\cite{WangS2013,BansNL2016}, over an order of magnitude larger than here. However, we note that $l_{m}$ is not a limiting factor at RT, where $\mu$ is limited by electron-phonon scattering\cite{HwanPRB2008,ChenNN2008,ParkNL14}.

In order to investigate the effect of grain boundaries on $\mu$, we study $\mu_{FE}$ of Poly-SLG, CVD-grown, encapsulated in hBN by using the same approach. Due to the small domain size, multiple crystal orientation are present within the same Hall bar. Poly-SLG is grown on 35$\mu$m Cu, following Ref.\citenum{LiS2009}. The substrate is annealed in 20sccm H$_2$ up to 1000$^{\circ}$C for 30min. Then, 5 sccm of CH$_4$ is added for the growth to take place for an additional 30 min. The sample is then cooled in vacuum ($\sim$1.3 mbar) to RT and unloaded from the chamber. SLG is wet transferred onto Si+285nm SiO$_{2}$ and then encapsulated in hBN, following the same procedure as above, but with $T$=180$^{\circ}$C in the pick up step, as higher $T$ is required to pick Poly-SLG from SiO$_2$.

The Raman spectra at each stage of the transfer process are shown in Fig.\ref{fig:Fig7}. Compared to SC-SLG, the Raman parameters are more scattered, $\emph{e.g.}$ the variation of A(2D)/A(G) and Pos(G) is twice that of SC-SLG, indicating inhomogeneous doping. The mean I(2D)/I(G) and A(2D)/A(G) ratio are indeed $\sim$7.2$\pm$2.1 and $\sim$11.1$\pm$4.1, respectively, which indicate a mean doping $\sim$100meV $\pm$100meV. Pos(G) in unstrained graphene with such low doping is$\sim$1581.6cm$^{-1}$\cite{LeeNC2012}. In our samples, however, the mean Pos(G) is $\sim$1582.6cm$^{-1}\pm$0.9cm$^{-1}$, indicating the presence of  uniaxial(biaxial) strain between $\epsilon\sim$0.02\%(0.05\%) and $\epsilon \sim$0.03\%(0.09\%) \cite{MohiPRB2009,YoonPRL2011,ZabeNL2012}. The slope of $\Delta$Pos(2D)/$\Delta$Pos(G) in Fig.\ref{fig:Fig7}f is $\sim$1.5 lower than in SC-SLG, due to the inhomogeneous distribution of doping and strain in these samples.

The resistivity as a function of $V_{BG}$ is shown in Fig.\ref{fig:Fig8}. By performing a linear fit to the conductivity close to CNP at $V_{BG}\sim4V$, we get $\mu_{FE}=C_{ox}^{-1}\left(d\sigma\slash dV_{BG}\right)\sim7000cm^2V^{-1}s^{-1}$ and $5000cm^2V^{-1}s^{-1}$ for electrons and holes respectively at 290K. These values are one order of magnitude lower compared to SC-SLG, as expected from the presence of grain boundaries. The shoulder in the left hand side of the resistivity as a function of $V_{BG}$ at $V_{BG}\sim-8V$ suggests the presence of an additional CNP in the SLG channel, indicating inhomogeneous doping\cite{BlakSSC2009}, in good agreement with the Raman analysis.	

As $\mu$ in Poly-SLG is significantly lower that SC-SLG, we investigate the combined effect of thermal annealing and encapsulation in Poly-SLG. To do so, we first exfoliate hBN onto Si+285nm SiO$_2$. Poly-SLG is then transferred to Si+285nm SiO$_2$ using the polymer-based wet transfer detailed above\cite{LiS2009,BaeNN2010,BonaMT2012}. This results in SLG partially on SiO$_2$ and partially on exfoliated hBN. The samples are then annealed at 600$^{\circ}$C in Ar/H$_2$ (40/40sccm). Spatially-resolved Raman spectra are acquired before and after annealing, both on SLG on SiO$_2$ and on hBN. The annealing sharpens FWHM(2D) from a mean$\sim$27 to$\sim$23 after annealing, Fig.\ref{fig:Fig9}a. This suggests that annealing makes the sample more homogeneous in terms of doping/strain by aggregating adsorbates/contaminants in blisters as shown in Fig.\ref{fig:Fig9}b. The annealed SLG is then encapsulated by depositing a second hBN using a polymer stamp following Ref.\citenum{PizzNC2016}. Samples are then shaped into Hall bars and contacted using Ni/Pd electrodes (30/40nm). Fig.\ref{fig:Fig9}c plots the resistance as function of gate. Fig.\ref{fig:Fig9}d shows $\mu_{FE}$ in annealed and encapsulated Poly-SLG, indicating a peak $\mu_{FE}\sim30000cm^2V^{-1}s^{-1}$, showing that annealing improves $\mu_{FE}$ by a factor$\sim$5. We note that this is amongst the highest RT $\mu_{FE}$ claimed for Poly-SLG to date.

\section{\label{Concl}Conclusions}
We reported high mobility in CVD grown, single crystal SLG placed on Si+SiO$_2$ using wet transfer and subsequently encapsulated in hBN. By cleaning interfaces between SLG and hBN at$\sim180 ^{\circ}$C, we achieved mobilities up to up$\sim70000cm^2 V^{-1} s^{-1}$ at RT and$>120000cm^2 V^{-1} s^{-1}$ at 9K, comparable to values achieved via dry transfer. The samples show ballistic transport over$\sim$600nm at 9K. This confirms that encapsulation in hBN and interface cleaning enable high mobility even after conventional polymer-based wet transfer techniques. We also investigated the combined effect of annealing and encapsulation in polycrystalline SLG, achieving RT mobilities up to$\sim30000cm^2V^{-1}s^{-1}$. Our results show that a scalable process such as wet-transfer can be used for the assembly of high-performance devices where high mobility ($>10000 cm^2 V^{-1} s^{-1}$) is essential.
\section{Acknowledgements}
We acknowledge funding from the EU Graphene Flagship, ERC Grant Hetero2D, EPSRC Grants EP/K016636/1, EP/K01711X/1, EP/K017144/1, EP/N010345/1, and EP/L016087/1, Wolfson College Cambridge, the Elemental Strategy Initiative conducted by MEXT and JSPS KAKENHI Grant JP15K21722.

\end{document}